# Big Data Strategies for Data Center Infrastructure Management Using a 3D Gaming Platform

Matthew Hubbell, Andrew Moran, William Arcand, David Bestor, Bill Bergeron, Chansup Byun, Vijay Gadepally, Peter Michaleas, Julie Mullen, Andrew Prout, Albert Reuther, Antonio Rosa, Charles Yee, Jeremy Kepner

MIT Lincoln Laboratory, 244 Wood St., Lexington, MA 02420

*Abstract* – High Performance Computing (HPC) is intrinsically linked to effective Data Center Infrastructure Management (DCIM). Cloud services and HPC have become key components in Department of Defense and corporate Information Technology competitive strategies in the global and commercial spaces. As a result, the reliance on consistent, reliable Data Center space is more critical than ever. The costs and complexity of providing quality DCIM are constantly being tested and evaluated by the United States Government and companies such as Google, Microsoft and Facebook. This paper will demonstrate a system where Big Data strategies and 3D gaming technology is leveraged to successfully monitor and analyze multiple HPC systems and a lights-out modular HP EcoPOD 240a Data Center on a singular platform. Big Data technology and a 3D gaming platform enables the relative real time monitoring of 5000 environmental sensors, more than 3500 IT data points and display visual analytics of the overall operating condition of the Data Center from a command center over 100 miles away. In addition, the Big Data model allows for in depth analysis of historical trends and conditions to optimize operations achieving even greater efficiencies and reliability.

## I. Introduction

Data Center Infrastructure Management (DCIM) is the practice of combining the management of the IT systems with the facility management systems. Gartner defines the goal of the industry as "integrating facets of system management with building management and energy management, with a focus on IT assets and the physical infrastructure needed to support them" [1]. Traditionally these roles were relatively independent activities. The IT groups concerned themselves with the servers, storage, and networks while the facility or building management departments were responsible for the data center.

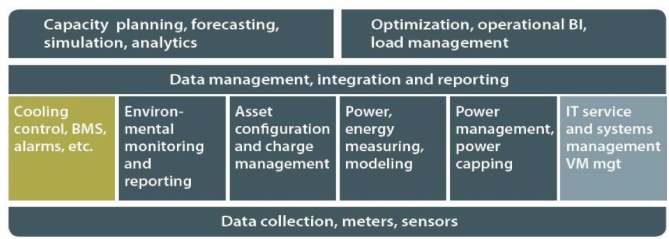

Figure 1. Representative model of DCIM and the underlying components from 451 Group and Forrester Research [3]

It is estimated by 2016, 70% of data centers will be utilizing a DCIM solution representing a $1.7B market [2]. Figure 1 depicts a representative model of DCIM and the underlying components that the 451 Group and Forrester Research built. They highlight the inputs to the system, their influence within the system, and the reporting and analytic tools expected of the operators.

As HPC and cloud services expand, an organization's ability to manage energy costs and fulfill the growing need of data center space is no longer easily sustainable. This is accompanied by the heavy demand HPC places on the data center. The need for cooling and power can vary greatly from periods of high usage to inactivity. The stability of the HPC cluster data center environment is as important now as the clusters themselves. The data center needs to be able to adapt to the variability in the overall environment as well as communicate to the HPC administrators of a potential problem that could pose a risk to the cluster. An organization's investment in HPC or cloud services infrastructure can range from $1 million to well over $100 million. Treating the data center and the IT equipment as a singular eco-system is essential to successfully leveraging the capability these systems have to deliver.

Traditional tool sets for managing data centers are not unlike those of other complicated integrated systems. The controls systems, environmental sensors, HVAC systems, and access control system are often managed from independent interfaces of the vendor-supplied solution. The expectation of the operators is to gain insight from the data correlated across solutions and their interfaces to complete the picture of the overall health of their system. There are many vendors who attempt to aggregate some of these interfaces to common DCIM suites. Many of the most successful vendors in the DCIM space include CA Technologies, Emerson Network Power, Schneider Electric, Panduit, and Raritan. Depending on the provider, many of these suites are heavily oriented or biased to a particular facet. For example, the Emerson and Schneider Electric tools are heavily focused on providing details about the energy consumption, power usage effectiveness (PUE), carbon usage effectiveness (CUE), and overall energy profile. An alternative would be tools from CA and Raritan whose focus begins from the IT side and provides more detailed information about the CPU, disk health, and remote access capabilities. Many of the solutions, while providing graphically informative widgets and charts, are often lacking a robust data visualization or analytic capabilities to gain insight into the behavior of the system as a whole. The goal is a solution where the IT and the facilities environment converge on a single platform. This allows IT administrators and facilities personnel to view and interact with the system in totality and gain insight into how to properly manage the interdependent systems while enabling advanced analytic capabilities and historical data reconstruction.

## II. APPROACH

Developing a platform for a converged DCIM solution began with previous success in IT system data aggregation utilizing Big Data strategies and 3D gaming technology. The existing 3D gaming model of monitoring and system management of the MIT Lincoln Laboratory 8000 CPU TX-Green and 900 CPU TX-2500 on-demand interactive HPC systems using the Unity game engine was the initial building blocks for development [4,5,6].

The 3D monitoring and management (MM3D) tool developed at MIT Lincoln Laboratory aggregates the diverse data flows of server information, HPC scheduler data, network data, storage data and user information and generates 3D visual representation depicting real time activity. This provides the system administration team a singular platform to manage compute resources and identify system troubles. The platform is a multiplayer environment where system administrators can interact with the physical world through the virtual one and reboot, reimage, remove from the scheduler, or otherwise manage the system. (Figure 2)

The common platform to visualize and manage system components reduces the dependency on vendor-supplied interfaces whose lack of interconnectedness limits the ability to have a comprehensive knowledge of the systems status and health.

To accommodate growing HPC needs while limiting organizational costs to ensure the investments were focused on delivering the maximal HPC benefit, Lincoln Laboratory invested in the modular Hewlett Packard 240a EcoPOD. The 240a is one of the most efficient data center solutions available, delivering 1.5MW of compute capacity in 44 racks with an effective PUE of 1.05 to 1.20 [7]. The EcoPOD allows the organization to efficiently scale HPC capabilities maximizing the computational resources dedicated to scientific research as compared to a brick and mortar infrastructure.

Vertically integrating HPC infrastructure with IT server management without a significant increase in overhead required a unique approach to monitoring and managing the overall system. To accomplish this, one must view the data center and the computational resources as a singular entity, which combines power, cooling, network, processing, storage, and scheduling to deliver a computational capability to the researcher. The data to achieve this was farmed from a variety of sources at relatively real-time intervals: core IT – gather software image versions, overall health, CPU load, memory allocations, disk utilization; network – gather link utilization; storage – health and capacity status; scheduler – available job slots, whose jobs are scheduled on which nodes; environmental systems – temperature, humidity, air pressure; power systems – status, voltages, amperages, and distributed and total KW. The data aggregation represents over 5,000 environmental sensor data points as well as over 3,500 data points for server information, health, and status. When approaching the monitoring and management in this way the data profile resembles many of the Big Data problems the research community use HPC resources to solve. The data consists of many of the V's associated with Big Data, a variety of sources, available at a high velocity, and over time representing a large volume of data [8].

This Big Data problem was approached using the same tools utilized for Bioinformatics, Cyber Intelligence, and Advanced Analytic research: a combination of Accumulo, MATLAB, and Dynamically Distributed Dimensional Data Model (D4M) [9].

A robust DCIM solutions should always provide the organization or the stakeholders with a common platform from which to visualize and analyze the data coming from the Data Center, provide a mechanism for historical data to be accessed, possess an alerting mechanism, inventory awareness, control, and provide insight into the interconnectedness of the Data Center system [10]. The process to get there is often difficult due to the disparate data sources and formats. Leveraging the D4M 2.0 data schema significantly reduces the complexity of handling the data by allowing the normalization of all the data to a simple tab separated .tsv file format. This allows for fast ingest into Accumulo, all the while normalizing the historical data records for future data analysis and removing the common complexity of managing data formats from multiple sources

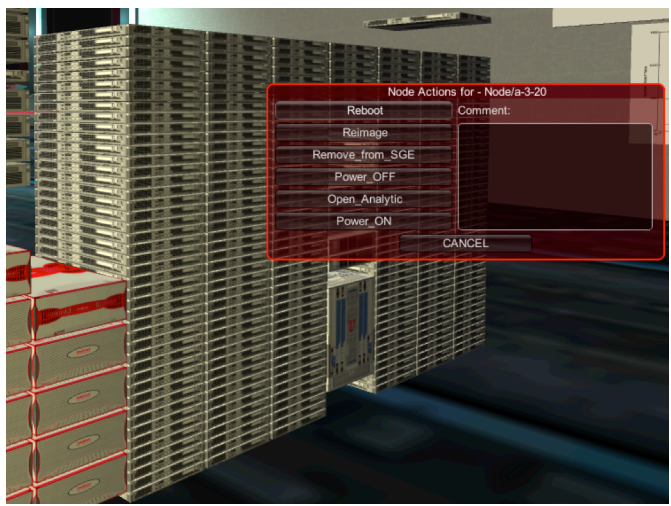

Figure 2. With a mouse up event you are able to bring up a menu to allow for in-game actions that will be taken in the physical world. The system allows for the user to add a comment, and it will logs the event and the state of the node at the time of the event for auditing and security purposes.

[11]. The 2.0 schema leverages the open schema concepts available in non-SQL based databases like Accumulo and allows for an open table structure. The core strength of this approach is the flexibility to add sensors and new monitoring capabilities without concern to the underlying data structure. This approach in a traditional SQL based system would prove to be very limiting, disruptive and unsustainable in the long term.

## III. DATA FLOW

The data for the DCIM implementation originates from two compute cluster environments in two data centers over 100 miles apart. Data collectors are run in the form of shell scripts executed through cron to gather node data, memory allocation, CPU load, image version, and scheduler information from both systems. The data is parsed and ingested into an Accumulo table. Simultaneously the Modbus registers embedded in the EcoPod are polled to gather environmental data. The ModBus registers include data about temperature, humidity, airflow, power utilization, air pressure, overall health and status. Data collection is achieved through a persistent C program leveraging libmodbus v3.0.4 which allows for the polling of the TCP/IP registers in bit fields as well as the real integer data formats. The data is then processed through the MATLAB program to generate a data input file or URI (universal resource identifier) that is ingested into the Unity3D game server.

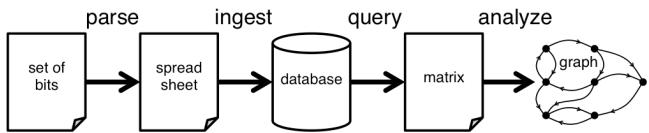

Figure 3. The standard steps in a data processing system to build towards a data analytic platform. (8)

As an example, the data model for processing the environmental conditions in the EcoPOD requires four core steps as shown in Figure 3; polling and parsing the registers for data, ingest the data into the database, query and perform correlation analysis to identify deviations in state, and finally generate the visualization input file for the 3D interface. Figure 4 shows the time necessary to perform each step for processing the environmental sensors available in the EcoPOD. The registers in the EcoPOD represent 5,325 unique data points and only take 5.2 seconds to gather, this is referenced against the baseline configuration file to check for deviations in state or the presence of alarms using matrix multiply operations in 1.06 seconds. It only takes 4.4 seconds to then ingest the results into Accumulo for future use and the visualization file is finally generated in 1.1 seconds. The total data processing pipeline is executed in 11.9 seconds.

The HPC monitoring and management component is derived from a series of network collectors running on each compute node. Critical information is gathered about the CPU load, memory allocations, disk information, software version, kernel version, IP, and MAC addresses. The relevant node data is ingested into Accumulo tables via Matlab/D4M and correlated against a baseline configuration file which defines inventory, expected configurations, and set alert values for critical system components. The baseline configuration file allows for easy changes to expected system profile as well as inventory additions or subtractions. Based upon system values the compute environments are virtualized in 3D accurately reflecting their physical layout for easy identification. This enables the fastest time to insight to actionable system status information for the administration team. For example, in Figure 5, one is able to identify from a single view, the systems currently scheduled with jobs, which machines have exhausted their jobs slots, and which machines have high loads. They are able to identify machines, which are in need of reimaging or are no longer accessing the file system and need attention. They are also able to use in-game tools to segregate machines, users, jobs, or CPU load to further investigate the state of the system.

An authoritative game server model enables the distribution of updated information to all connected clients/users via a series of remote procedure calls to ensure data integrity across all connected platforms. The authoritative model also allows for access controls to be in place for tiered privileges to view state or take certain actions against machines. Using Unity in an authoritative configuration makes for a platform agnostic environment where Windows, MAC, Linux, iOS, Android, and Blackberry are supported simultaneously.

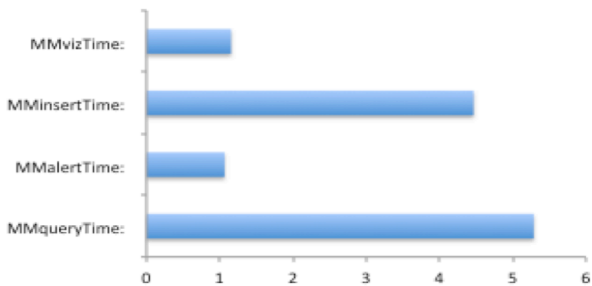

Figure 4. Time in seconds required for data processing, analysis, database insert and build the visualization configuration file.

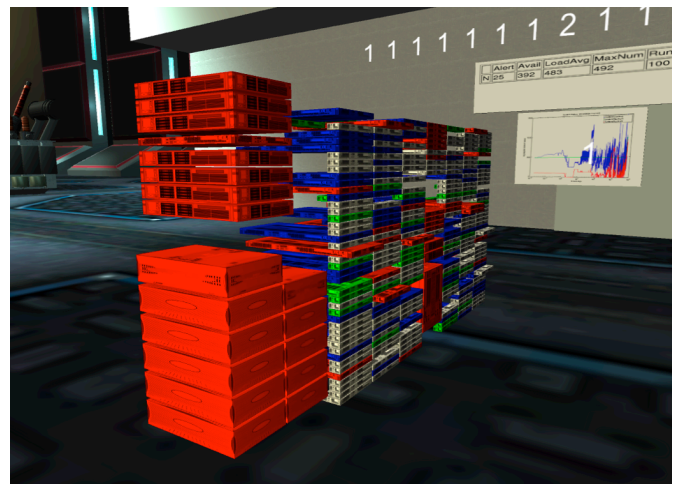

Figure 5: The image shows the cluster view of different visual cues used to depict in game alerts or status.

## IV. RESULTS

The converged DCIM tool implemented at Lincoln Laboratory has proven to be a very effective tool to rapidly gain situational awareness of a complex and varied environment. The initial configuration requires the most hands-on involvement to establish the assets of interest. A baseline of expected operating conditions is established to identify trigger points for visualizations and alerting mechanisms. Deviations from the baseline will trigger MIN, MAX or binary status alerts in the form of visualizations in 3D space as well as email alerts depending upon the severity of the trigger. Informational status is displayed to the user by taking advantage of the flexibility provided in 3D space. The state of current operating conditions is graphically represented in realtime in a way that the administrator or facilities operator can easily interpret.

Care was taken to use imagery and visual cues, which are commonly associated with events or deliberate in their definition, rather than establish completely original graphical imagery. This decision allows the operator or end user a faster time to process a commonly accepted visual cue and remove the barrier of learning a new library of graphics. (Too often vendors want to differentiate their product and establish a unique set of imagery leading to confusing menus and definitions of graphical choices.)

In Figure 6a, informational status is represented via easy to understand snowflake icon to indicate mechanical cooling is in use. The air propagating up the outside of the EcoPOD indicates the outside dampers are open and taking advantage of the outside air and are operating in economizer mode. Water events such as humidity, float sensor alarms, dew point levels are easily identified by an unmistakable animation of water originating from the affected component. Important electrical thresholds such as total IT power usage is easily indicated through a lighting bolt emanating from the affected power feed as shown in Figure 6b. This is repeated with temperature alerts as well as more severe events such as a fire. It should be noted that this system is not used as a substitute for traditional life safety mechanisms and is a situational awareness tool. In the event of a fire or severe failure, traditional alarms are connected to core critical alarm control infrastructure to alert the appropriate authorities and personnel.

The 3D gaming model of IT system monitoring and management has shown to be successful in relaying accurate system state in a rapid, easy-to-digest platform. In Figure 5 one can see the TX-Green super computer system in its actual rack configuration allowing for easy asset identification for the system administration team. The systems are color coded and sized according to system status. A colorless node indicates its status is healthy, idle, and awaiting job submission. Nodes represented as blue are actively running jobs and have more than half of their computational cores scheduled. A green node is actively running user submitted jobs; however, fewer than half of the cores are allocated. A red node indicates it has failed a system check. This can be a node, which is out of sync with the common image, has exhausted a memory threshold, or has a failed component. These machines should be addressed using the appropriate in-game action. The Y scale of the node identifies the relative CPU load giving an indication of computational activity. Any single indicator does not necessarily trigger an action; it is the ability to quickly assess and identify abnormal system behavior, which is most valuable.

A common example of functional use is when a user's job is spread over 128 machines and has experienced a memory leak. This leak has exhausted the available memory across the nodes in use, causing most of the machines to stop responding and turn red within the game, thereby triggering an admin to investigate further. The admin will seek to identify all the affected nodes and the nodes, which have not yet exhausted their memory. These nodes will be aggregated using an in-game "pull" command to query the user's name or job number to isolate the nodes where the job is scheduled and take the appropriate action to bring them back into service. By enabling the administration team to easily identify the other jobs scheduled on those nodes, they are able to notify the researchers to resubmit their jobs or alert them to the possibility of their computation not finishing. The isolation of these events also helps identify who in the user community may be in need

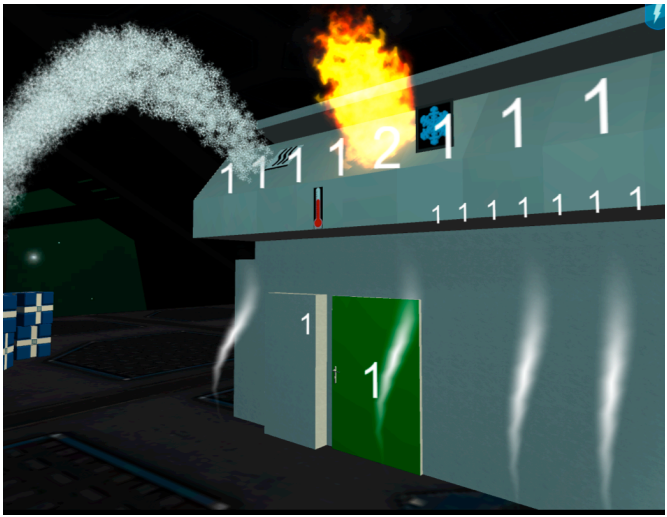

Figure 6a - The images show an arrangement of different visual cues used to depict ingame alerts or status.

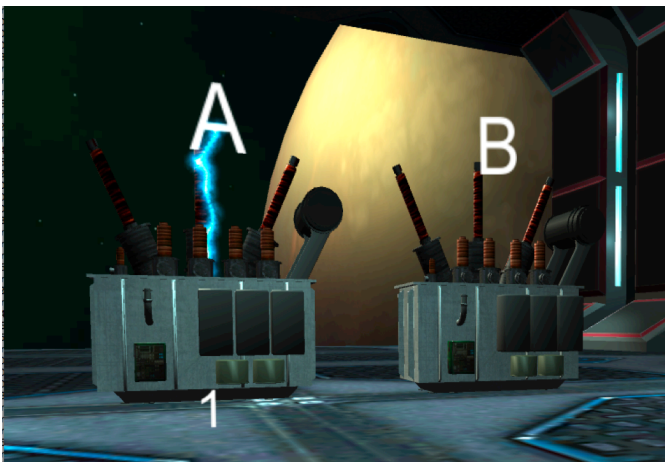

Figure 6b. An example of an electrical event to power Feed A of further consultation to optimize their code, bolstering the effectiveness of user support services.

## V. 'SERIOUS GAMES' MODEL

The concept of using games for serious work is not new and can be credited to Clark Abt who championed the use of games, albeit board games, in his 1970 book *Serious Games* [12]. In his work he defended the use of games for educational, training, or business purposes and insisted the application of serious concepts to games should still be grounded in entertainment. The 3D gaming platform for implementing a DCIM solution has been successful for multiple reasons. First, the model has served well for the millions of gamers who have self-identified the platform as an acceptable interface to convey interactive actionable data. Gamers spend countless hours interacting and responding to in-game information, and gaming environments have shown to be a far more engaging and rich vehicle to convey information than traditional web platforms. The platform requires little training to work within as most people have spent a substantial amount time using 3D environments for personal entertainment, ultimately reducing the time and complexity of learning a new tool. Argwal identified "ease of use" as the critical component to cognitive absorption of information and use of IT tools [13]. The end user is not threatened by a game and can be confident in their ability to navigate a familiar set of controls thereby furthering the likelihood of adoption and reducing user fatigue.

The flexibility provided by Unity for platform independence allows the building of game binaries for all major platforms of desktop computers as well as mobile devices. This enables the administration team to have quick, easy access to system status and health at all times. The gaming platform also provides a flexible multiplayer/user environment where team members or managers can log into a common environment and work collaboratively in a virtual space.

Lastly, the 3D environment allows for a more diverse tool set to cue the operators' senses to convey alerts or critical information: for example, using colors, shapes, and morphing of objects as a means to attract the attention of the administration team to a change in the environment. Using Xbox controllers the vibration of the controller is used to alert the administrator to a new event or to positively reinforce in game actions by providing feedback through touch. The 3D gaming environment also allows for seamless integration of in-game sounds. The combination of these tools enable the targeting of three of the five core senses greatly increasing the cognitive impact of alerting thereby reducing operator fatigue.

The real time operational Big Data 3D DCIM tool has proven to be very effective for alerting, but it is the common platform that provides the next layer of insight often difficult to attain. For example, this tool enables the end user to visualize the correlation of users' processing needs and the impact to the operating profile of the EcoPOD in connection to the external environment. When power alerts are observed for a spike in KW usage, it can be directly correlated to the initiation of a large job a user may have just launched. The job may run for a few hours and cause the temperature to slowly rise to the point where mechanical cooling needs to be activated ultimately reducing the overall PUE. By visualizing the content of the Data Center and its components behaving as a system and responding to real conditions, the operator is able to gain insight into the totality and interconnectedness that is lost when viewing the components as stovepiped services.

Understanding the system and its impact on resources, one is better able to identify the capacity needs for IT, power, cooling, and space. One example is being able to identify hot spots and using the scheduler to distribute users tasks across racks to take pressure off the cooling systems during the summer months thereby reducing the need to engage mechanical cooling. The granular level of insight into the specific micro climates, which exist in the Data Center, allows the administrative user to have an appreciation of space planning and improved understanding of the different cooling needs of storage equipment versus compute and networking equipment. The impact to capacity planning is felt when planning to add assets to the Data Center; this helps identify the best locations to take advantage of cooling and available power.

To this point this paper has focused on the real time insight the converged DCIM tool provides in daily operations. However, one of the most powerful components this model enables is a deep historical analytic capability. The data collected from the IT infrastructure and the environmental sensors are all stored in an Apache Accumulo table, which is based upon the Big Table design and sits on top of Hadoop [14]. As was highlighted in the data pipeline, D4M is used to ingest and query the data from the database. The combination of Accumulo and D4M allows for the data to be handled using mathematical expressions. "By leveraging the mathematical abstraction across all steps, the construction time of a data processing system can be reduced." [15] The efficiency with which historical analytics can be run and developed is a key tenet of this platform.

The current platform is designed across two databases for a two year period consists of over 15 billion entries providing a constructible history of job submission, machine health, and data center operating conditions. This enables the development of advanced analytics to profile the system for reporting, planning, and operational support. Some examples are user job submissions to illustrate user metrics and understand how users are interacting with the system. Identifying periods of the year where usage is heavy or understanding that Friday is a very popular day to submit long jobs for the weekend resulting in a spike of storage utilization and increased power consumption are valuable. Further, reviewing machine utilization, identifying hardware failures to help with inventory control for spare parts, and identifying components' performance across multiple platforms are also quite valuable. Most important is the ability to revisit and play back the exact conditions which existed before, during, and after a real event such as a smoke alarm or high temperature warnings. This has enabled the identification of patterns in the humidity and airflow data, which caused the system to become erratic. This then prompted the identification and reconfiguration of panels and space allocation to account for an unbalanced load. One can reconstruct the psychometric chart and fully grasp the operating profile at the specific time. Without the forensic capability one would only speculate on what could cause the differential pressure to suddenly drop in certain areas of the data center and never fully appreciate the relationship of the components involved in the root cause.

## VI. Conclusion and Future Work

The benefits of leveraging Big Data and the 3D gaming platform for a converged DCIM solution have been established. The product of the research meets the key requirements set forth in the industry: that a DCIM tool should provide the stakeholders with real time insight into critical facilities systems, IT infrastructure, life cycle management, cost reduction strategies, historical analytic capabilities with less effort, less cost, and greater capabilities than the traditional commercial offerings.

There is still more work to be done in developing the capability further by creating a smart tool which can help take advantage of green opportunities. For example, one could be to integrate with the scheduler to delay jobs until the evening hours or off peak hours to take advantage of lower electric rates. The analytic capability exists to profile jobs prior to launching, so that during the peak summer months the longer, computationally-intensive jobs could be queued to run during periods of lower temperatures in early morning hours or in the evenings depending on users priorities. Other future work includes improving the visualizations and internal game play to take advantages of gamification strategies to prolong interaction with the tool as well as incentivize positive in-game behavior.